\documentclass[%
 reprint,
 amsmath,amssymb,
 aps
]{revtex4-2}

\usepackage{graphicx}
\usepackage{dcolumn}
\usepackage{bm}
\usepackage{subfigure}
\usepackage{tikz-cd}

\begin{document}

\preprint{APS/123-QED}

\title{Robust methods to detect coupling\\ among nonlinear dynamic time series}

\author{Timothy Sauer}
\affiliation{%
George Mason University\\ Fairfax, VA 22030
}%
%
%
\author{George Sugihara}
\affiliation{
Scripps Institution of Oceanography at UCSD\\La Jolla, CA 92037
}%
%
%
\date{\today}

\begin{abstract}
Two numerical methods are proposed to detect and analyze coupling in time series from deterministic nonlinear systems. The first identifies the presence of coupling or interdependence, while the second determines the directionality of coupling. The second method can also identify latent coupling - series that are not directly coupled but are correlated due to influence by another, unobserved system. Both methods accommodate periodic, aperiodic, and chaotic dynamics, and use order statistics derived from relative distances within a time-delay embedding. The methods are intended to be practical and apply to data sets consisting of simultaneous system recordings, and robust to observational noise due to their use of order statistics.

\end{abstract}

\maketitle

\section{Introduction}
Determining causal relationships from time series data is a core challenge in many scientific and engineering fields. This task becomes especially difficult when dealing with nonlinear dynamic systems, where the deterministic effects of one component on another may be indirect, delayed, or obscured. In such settings, the assumption that causes act independently often fails due to feedback and coupling. More importantly, with nonlinear dynamics, which are ubiquitous in nature, causal relationships can occur even when there is no measurable correlation between variables or coupled systems\cite{Sugihara2012detecting}.

The traditional link between causation and correlation reflects the historical influence of linear models, where Pearson correlation is often taken as a necessary condition for causal influence. But this association breaks down in nonlinear systems. Even simple examples, such as the two-species logistic model \cite{Sugihara2012detecting}, demonstrate that one system can drive another without showing any observable correlation.  Given the widespread nonlinearity observed in natural systems such cases undermine widely used classical frameworks such as -- path analysis \cite{wright1934method}, structural equation modeling \cite{bollen1989structural}, and causal graphs \cite{pearl2009models} among them -- which all require the presence of correlation as preliminary to infer causation.

Over the past two decades, several techniques have been developed to infer causality in nonlinear systems, including methods based on state-space reconstruction. These approaches have been effective in identifying whether the dynamics of one system influence another, a concept often referred to as dynamic causation. However, distinguishing between different types of coupling -- unidirectional, bidirectional, or driven by an unobserved third system -- remains a significant challenge, particularly in noisy or short data sets.

In this paper, we introduce two numerical methods to detect and characterize causal coupling in time series from nonlinear deterministic systems. The first method tests for the presence of coupling, while the second identifies its direction and can reveal cases of latent coupling caused by an unobserved driver. Both rely on order statistics derived from relative distances in a reconstructed phase space, allowing the methods to handle periodic, aperiodic, and chaotic dynamics. Because the analysis depends only on the ranks of distances -- not their magnitudes -- these methods are robust to observational noise and well-suited for practical use in real-world datasets.

 Given two nonlinear dynamical systems $U$ and $V$ whose dynamics are asymptotic to a finite-dimensional attractor, and their simultaneous time series observations $\{u_t\}$ and $\{v_t\}$, we develop criteria to decide which of the following relations hold:

\begin{enumerate}
    \item \textbf{Unidirectional Coupling:} U drives V (\(U \to V\)) or V drives U (\(V \to U\)).
    \item \textbf{Bidirectional Coupling/GS:} Both U and V drive each other (\(U \leftrightarrow V\)), or the systems are in generalized synchrony (GS).
    \item \textbf{Latent Coupling:} U and V do not influence each other but appear coupled because they are driven by a third, unobserved dynamical system.
    \item \textbf{Independence:} U and V are dynamically independent of one another.
\end{enumerate}
 
 Our methods rely on {\it genericity } of the dynamics, in that counterexamples to these criteria may be artificially constructed, but are not expected to be common in general circumstances.

{\it Generalized synchrony} (GS) \cite{rulkov1995generalized,pecora2000detecting}  is the circumstance when each state of $U$ occurs simultaneously with a unique corresponding state of $V$, and vice versa. Otherwise said, there is a one-to-one invertible function from the states of $U$ to the states of $V$. This is a generalization of the special case of {\it identical synchrony} \cite{pecora1990synchronization} where $U=V$ and the corresponding states are identical by defintion. Generalized synchrony is often observed as a result of unidirectional driving that is relatively strong, although (perhaps counterintuitively) in specific examples it can also occur for moderate unidirectional driving. An example of this, and a more detailed description of GS, is contained in Section \ref{sGS}. 

In systems that are dominated by linear dynamics or noise, Granger causality \cite{granger1969investigating} is recognized as an effective means of ascertaining the direction of coupling from observed time series. Granger causality is based on the intuitive idea that causal coupling $U\to V$ is established when the prediction of system $V$ improves when observations from system $U$ are included in the model. Implementations using linear cointegration, entropy \cite{schreiber2000measuring} or mutual information \cite{sun2014causation,runge2018causal} have been developed and used successfully for this purpose.  

Therefore, it is somewhat surprising that for finite-dimensional, deterministic nonlinear systems under generic conditions, Granger causality-based methods are ill-posed. Indeed, the phenomenon of generalized synchrony means that if $U$ drives $V$, once the state of $V$ is identified from observations, $U$ obviously can provide no further information about the state of $V$, meaning that Granger causality has no power to identify the fact that $U\to V$. Moreover, even without synchrony, Takens' theorem  \cite{Takens,SYC} shows that in the large data limit, $U$ cannot add more information because the state of $U$ can be determined entirely from histories of $V$. Conceptually, the lack of separability of causes that lies at the foundation of Granger causality undermines its theoretical basis for making accurate directionality assessments in deterministic nonlinear systems (see \cite{Sugihara2012detecting,cummins2015efficacy} for further discussion of this issue). 

The concepts of time-delay embedding imply that each state of a dynamical system can be represented uniquely by a sufficiently long delay vector constructed from time series observations of that system. Assume that $U$ and $V$ are deterministic dynamical systems and that $U$ has an input to $V$, but that there is no return input from $V$ incident on $U$. Then, under technical conditions that are satisfied in general circumstances \cite{Takens,SYC,deyle2011generalized}, we can expect to reconstruct the entire combined dynamical system $U\to V$ from observations $v_t$ of $V$. The same is not possible from observations on $U$ only, because observations on $U$ will lack information about $V$, and therefore fail to reconstruct the entire coupled system. We exploit this asymmetry in our methods.

This asymmetry has been exploited before, for example in \cite{schiff1996detecting,arnhold1999robust,osterhage2007measuring}, and more broadly in the Convergent Cross Mapping (CCM) method \cite{Sugihara2012detecting}. In \cite{Sugihara2012detecting}, to distinguish time series data generated as projections from an attractor from statistical phenomena, CCM includes the crucial ingredient of tracking the asymmetry as a function of data length, which has been widely successfully when applied to experimental and field data \cite{ye2015distinguishing,gao2023causal}. The continuity statistic of Pecora et al.~\cite{pecora1995statistics} is also based on this asymmetry. Other methods relying on state space reconstruction to detect coupling include \cite{harnack2017topological,leng2020partial}, and a version based on sorting in \cite{breston2021convergent}. 

 Exploitation of asymmetry of state reconstruction in the latent coupling case was previously addressed for deterministic systems by \cite{sauer2004reconstruction}. This work demonstrated in theory why it should be possible to reconstruct the unobserved driver in latent coupling. More recently, this theoretical idea was developed into a readily implementable method \cite{gilpin2025recurrences}. 

Inspired by these prior results, in this article we describe a unifying approach that exploits the asymmetry in delay-coordinate reconstructions to affirmatively detect directional and latent coupling. The use of order statistics is intended to minimize the  requirement on data length, thereby making it more broadly applicable to real-world problems often constrained by limited data availability. The approach is robust in that it depends only the relative pairwise distances of points in the reconstructions. 

\section{Delay embedding}
The tests to be described in section III rely on the reconstruction of attractor dynamics from time series observations. The idea was described early on by Crutchfield et al. \cite{crutchfield1979,packard1980geometry} and formalized in the embedding theorem of Takens \cite{Takens}. This foundational work led to applications in a wide array of fields, including geophysics \cite{hossain2024,astudillo2010}, neuroscience \cite{kwessi2021,schiff1996detecting,arnhold1999robust,osterhage2007measuring}, ecology \cite{Sugihara2012detecting,deyle2016tracking,munch2023constraining},
fault analysis \cite{lee2023}, and data assimilation \cite{hamilton2016ensemble}.

Assume \(\{u_t\}\), \(1 \leq t \leq N\), is a time series sampled at discrete intervals from a finite-dimensional, compact attractor \(U\) of a dynamical system. At each time \(t\), define the observation vector 
\[
U_t = [u_t, u_{t-1}, \dots, u_{t-e+1}],
\]
where the integer \(e\) is the embedding dimension. The collection of these vectors, \(\{U_t\}\), forms the time-delay reconstruction of the attractor \(U\). Note that this process relies on a measurement function \(h_U : U \to \mathbb{R}\), which maps the system's state to a scalar observable, \(u_t = h(U(t))\).

The main theorem of \cite{Takens} states that for generic conditions and sufficiently large $e$, there is a one-to-one bijection between the states of a smooth manifold $U$ and the reconstructed states $U_t= [u_t, u_{t-1}, \ldots, u_{t-e+1}]$. The underlying genericity assumption for this theorem is that both the dynamics of $U$ and the observation function $h_U$ are not special, meaning that if it fails for $U$, there is another system infinitesimally close to $U$ for which the conclusion holds. Later work \cite{SYC} simplified the main idea somewhat, by including the case where the attracting set may not be a smooth manifold, and requiring genericity on the observation function only, assuming mild and specific conditions on the equilibria of the attractor $U$. The latter article also replaced the concept of {\it genericity}  with a more probabilistic concept of {\it prevalence}.

Expanded versions of Takens' theorem show that in place of univariate time series, multivariate time series may be used to reconstruct the dynamical system \cite{SYC,deyle2011generalized}.  In the present article, the results remain unchanged whether the reconstruction is made from univariate or multivariate times series, since only the fundamental property of a one-to-one correspondence is relevant to the success of the statistical tests. The major requirement we make is on genericity of the dynamics and coupling.  That is, for each state of $U$, or $U$ and $V$ if they are coupled, we assume that both the dynamics and the observations are not special, so that the delay coordinate reconstruction faithfully replicates the upstream dynamical states.

Extensions of embedding theorems for coupled systems as discussed in this article are proved in \cite{cummins2015efficacy}. This reference applies a method of \cite{pecora1995statistics} to assess certain coupling results from time series recordings.

\section{Statistical Tests}

In this section, we propose two statistical tests to analyze coupling. The first test (DetC) evaluates whether two systems are coupled at all. If this test is positive (i.e. relation 4 above is ruled out), the second test (DirC) is applied to determine the direction of influence or, more precisely, to distinguish among the relationships outlined in relations 1 - 3 above.
\subsection{Detection of coupling}

 Assume that two simultaneous time series $\{x_t\}$ and $\{y_t\}$ of length $N$ are measured from $U$ and $V$, respectively, and that  delay coordinate vectors $\{U_t\}$ and $\{V_t\}$ are formed as above.
Fix time $t$ and a number of nearest neighbors $n$. Using a Euclidean norm find the $n$ nearest neighbors $\{V_{t_1},\ldots, V_{t_n}\}$ of $V_t$. Consider the corresponding set $\{U_{t_1},\ldots, U_{t_n}\}$, representing the simultaneous states $U_{t_j}$ in the $U$ system, and examine their proximity to $U_t$, the simultaneous correlate to $V_t$.

If $U$ and $V$ are independent, the set $\{U_{t_1},\ldots, U_{t_n}\}$ should have no special relationship to $U_t$.
On the other hand, if $U$ drives $V$, then each $V_t$ determines a unique state in $U$ and $V$, and the nearest neighbors $V_{t_j}$ correspond to simultaneous $U_{t_j}$ that are relatively near $U_t$ (in particular much nearer than random choices from the $U$ time series). The same is true if instead of $U\to V$, there is a latent coupling, meaning a third system $D$ such that $D\to U$ and $D\to V$. 

This distinction can be analyzed statistically, due to the theory of order statistics (see, e.g., \cite{gentle2009computational}). Given a uniform random choice of $n$ numbers from the interval $[0,1]$, the $j$th smallest of the $n$ numbers follows a beta distribution Beta$(j,n-j+1)$, whose mean and variance  are $\mu=j/(n+1)$ and $\sigma^2=j(n+1-j)/((n+1)^2(n+2))$, respectively. 

We will apply this fact in the following way. Consider a time step $1\le t \le N$, the reconstructed state $V_t$, and as above, its $n$ nearest neighbors $V_{t_1},\ldots, V_{t_n}$ according to the Euclidean distance $d_i = |V_t - V_{t_i}|$ from $V_t$. Define the subset $U_{t_1},\ldots, U_{t_n}$ of contemporaneous delay vectors at the times $t_1, \ldots, t_n$, and their distances $d_i = |U_t - U_{t_i}|$ from $U_t$. As mentioned above, we would like to know whether the $U$ subset is as close to $U_t$ as the $V$ subset is close to $V_t$. 

Each distance $d_s = |U_t - U_s|$, $1 \le s \le N-e$, can be assigned a relative rank among all $N-e$ distances of the reconstructed states, as follows. If $d_s$ is the $R_s$th smallest distance from $U_s$, let $r_s = R_s/(N-e)$ denote the relative rank. By construction,  the $r_s$ are uniformly distributed between $0$ and $1$. If the subset $S = \{r_{s_1}, \ldots, r_{s_n}\}$ is randomly chosen from the entire set, order statistics gives the expected relative position of the $j$th smallest number $r_{(j)}$ of the subset to be $j/(n+1)$. On the other hand, if the subset $S$ is chosen randomly from a reduced proportion $p$ of the entire set, then the expected relative position will be $jp/(n+1)$. Therefore we can use 
\begin{equation} \label{e1}
\hat{p} = \frac{(n+1)r_{(j)}}{j}
\end{equation}
for each $1\le j \le n$, as an estimator for $p$. If the dynamical systems $U$ and $V$ are uncorrelated, we expect to recover $p = 1$ for each $j$. If there is a direct or indirect coupling between the two systems, the contemporaneous delay vectors in $S$ will be chosen from a portion of the attractor that is limited in extent, with a corresponding proportion of the state space $p<1$.

\begin{figure}
\subfigure[]{\includegraphics[width=.23\textwidth]{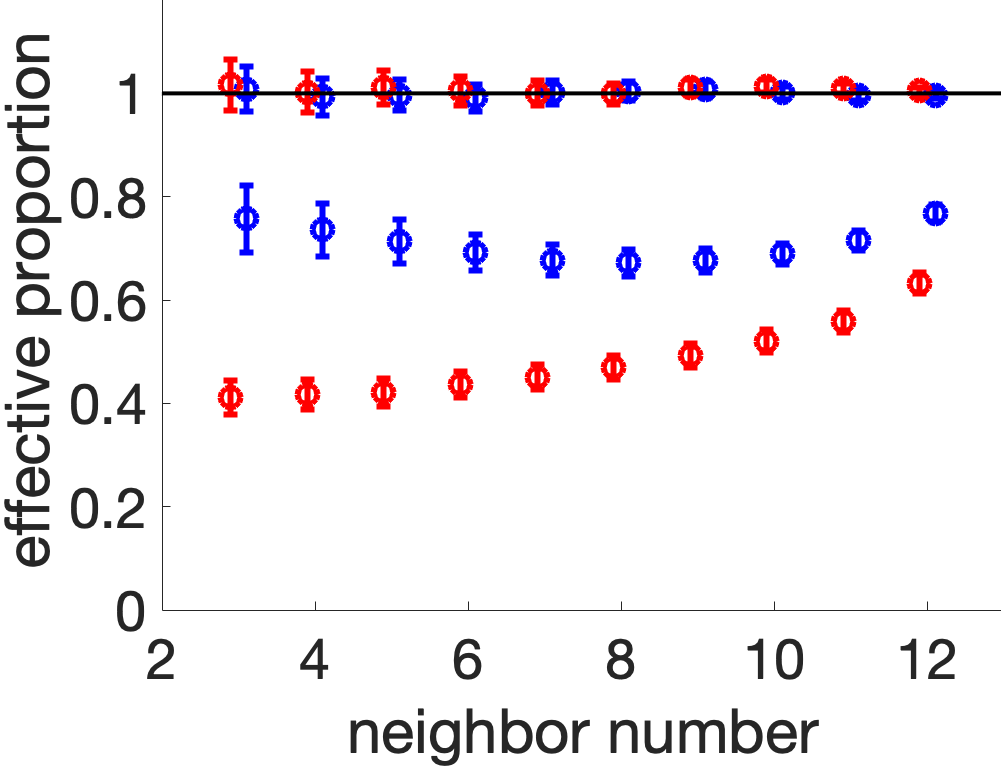}}\hspace*{.02in}
\subfigure[]{\includegraphics[width=.23\textwidth]{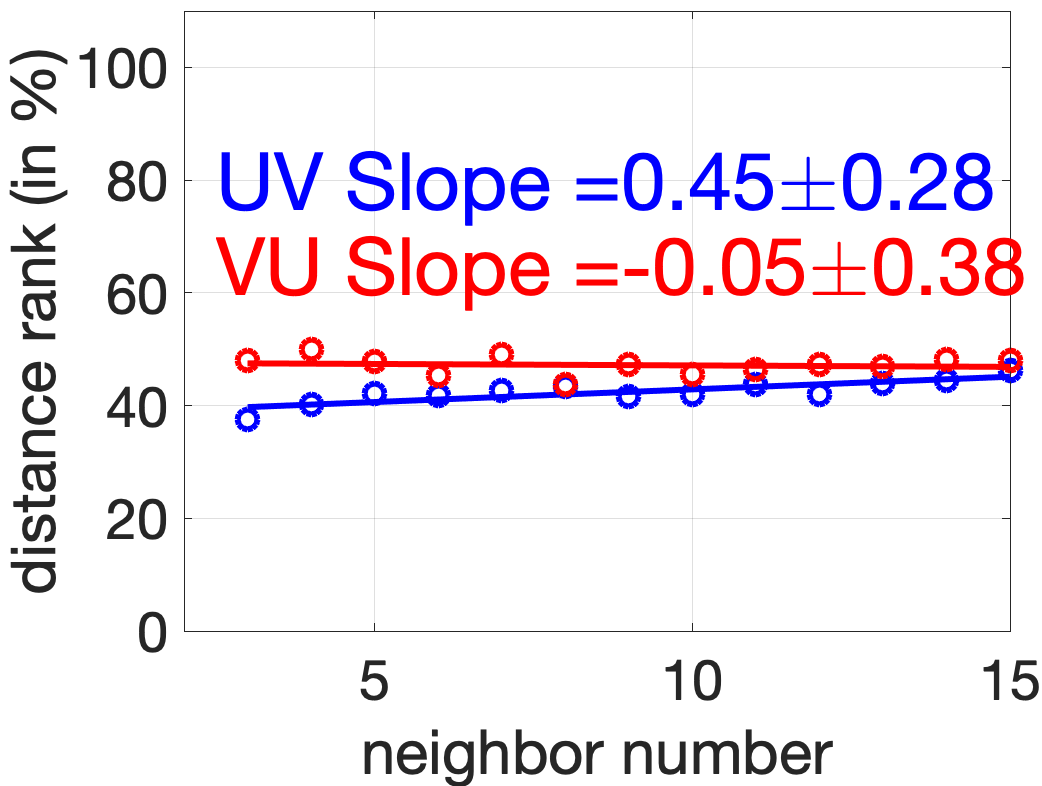}}
\caption{\label{f1} DetC  and DirC  tests applied to time series from skew system $U\to V$. Each of $U$ and $V$ are given by equations \eqref{e8}. (a) For zero forcing (upper curves), both DetC(U,V) (blue traces) and DetC(V,U) (red traces) show that $1$ lies inside the confidence intervals for $p$, and for positive forcing $c = 0.5$ (lower curves), $1$ lies outside the confidence intervals. (b) For $c = 0.5$, the slope estimate for DirC(U,V) (blue trace) rejects zero, and DirC(V,U) (red trace) does not, correctly implying a forcing $U\to V$.  }
\end{figure}

Summarizing this discussion, the following test averages the estimate over the time series $U_t$ as a null hypothesis for independence of the two time series. 

\medskip

\noindent {\bf Detection of Coupling Test: DetC(U,V).}  For each $t$, find the $n$ nearest neighbors of the delay vector $V_t$ and denote their times $t_1,\ldots t_n$. Sort the entire set of $N-e$ distances $|U_t - U_s|$ for $1\le s\ne t \le N-e+1$, and for each $1\le j \le n$, find the relative rank $r_{(j)}$ of the $j$th smallest distance in the subset $S = \{|U_t - U_{t_1}|,\ldots,|U_t - U_{t_n}|\}$, among the entire set $\{|U_t - U_{s}|\}$. For each $j$, $1\le j \le n$, an estimate $\hat{p}$ is found from the expected value  
\begin{equation}\label{e1a}
\hat{p} = \mathbb{E}\left[\frac{(n+1)r_{(j)}}{j}\right]
\end{equation}
averaged over $t$. The value of $p$ is equal to $1$ if and only if there is no coupling between $U$ and $V$.

\bigskip

A one-sample Student's t-test can be used to assign a confidence interval to the estimate $\hat{p}$ in \eqref{e1a}. A $95\%$ interval around  $\hat{p}$ contains the possible values of $p$ with $95\%$ certainty, and we would like to know whether the interval contains $p=1$. Since the one-sample t-distribution has $N-2$ degrees of freedom, it is well approximated by the normal distribution for large $N$, which represents two standard deviations for the $95\%$ level.  (Note that we can further average over $e$ to decrease the variance even more, although we have not done so in the examples, where a fixed $e = 8$ was used.)

Figure \ref{f1}(a) shows the results of the DetC test for observations of $U$ and $V$ where $U\to V$ for two different coupling strengths. When the coupling is zero (top red and blue curves represent DetC(U,V) and DetC(V,U), respectively) the null hypotheses $p=1$ is not rejected, and we conclude there is no coupling. When the coupling is positive (lower red and blue curves) the null is rejected, and coupling is concluded.

The length-1000 time series used to make Fig.~\ref{f1}(a) were generated by coupled discrete dynamical systems. Both $U$ and $V$ are  two-cell networks of H\'enon-like \cite{henon1976two} maps. Specifically, each $U$ or $V$ consists a four-dimensional discrete map of two H\'enon-like maps coupled together, with equations
\begin{eqnarray} \label{e8}
 x_{i+1} &=& b_1\cos x_i+c_1y_i+d_1z_i\nonumber\\
 y_{i+1} &=& x_i\nonumber\\
 z_{i+1} &=& b_2\cos z_i+c_2w_i+d_2x_i \\
 w_{i+1}&=& z_i \nonumber
 \end{eqnarray}
 Here the parameters $b_i, c_i$, and $d_i$ were chosen randomly near $2.2$, $0.1$, and $0.1$, respectively, which resulted in chaotic dynamics in the network. The time series observation from each network is the $x$ variable. The coupling between the two networks is achieved by adding the $x$-variable from system $U$ to the $z$-variable of system $V$, multiplied by a coupling strength (either $0$ or $0.5$)  in Fig.~\ref{f1}. 

The map in H\'enon's original paper \cite{henon1976two} is often studied due to the fact that it has a chaotic attractor near a basin boundary.
Here we use a variant of the original H\'enon map. The version used in (\ref{e8}) replaces a quadratic term in the $x$ equation with a cosine, which retains the chaotic dynamics but moves the basin boundary of the attractor far away, to enhance stability when used in a network. In equation (\ref{e8}), two such maps are coupled together to produce a hyperchaotic attractor of correlation dimension slightly greater than $2.0$. Each of the attractors $U$ and $V$ in Fig.~\ref{f1} and \ref{f2} is an attractor of this type.

\subsection{Direction of coupling}
The second test assesses directionality of the coupling. Our general assumption is that there is an unknown finite-dimensional attractor embodying one of the above joint dynamics scenarios 1 - 3,  and that it is observed by two time series that have each been used to reconstruct dynamics as in Takens' theorem. 

First assume that relation 1 holds. Thus a unidirectional coupling exists, which we will suppose, without loss of generality, is $U\to V$. Further assume that time series recorded from both $U$ and $V$ are used for time delay reconstruction using generic observations functions and sufficiently large embedding dimension $e$ required by Takens' Theorem. Let $R(U)$ (resp. $R(V)$) denote the reconstructed attractors, and let $T_{U,U}$, $T_{V,U}$ and $T_{V,V}$ denote the projections from the reconstructions to $U$ and $V$ as shown in the diagram:
\begin{center}
\begin{tikzcd}[column sep=small]
 R(U) \arrow[dr,swap,"T_{U,U}"] & &R(V)\arrow[dl,"T_{V,U}"] \arrow[dr,"T_{V,V}"]& \\
 & U & & V 
\end{tikzcd}
\end{center}
Takens' theorem implies that generically, there is a homeomorphism between $R(U)$ and $U$, which we denote here by $T_{U,U}$, and also between $R(V)$ and the skew product $U\to V$, which can be further projected to the two components by $T_{V,U}$ and $T_{V,V}$.  While the map $T_{U,U}$ is one-to-one, the other two maps are not: Although $R(V)$ reconstructs the skew product $U\to V$, unless $U$ and $V$ are in general synchronization, $T_{V,U}$ and $T_{V,V}$ will fail to be one-to-one, because there is no return coupling $V\to U$. In particular, the inverse image $T_{V,U}^{-1}(x)$ will generally be more than one point, since the state of $U$ does not determine the contemporaneous state of $V$.

As observers, our access to this diagram is only along the top row. According to the diagram, the reconstructed state $V_t$ in $R(V)$ corresponds to a single point $T_{U,U}^{-1} T_{V,U}\, V_t$ in $R(U)$. That means for each $V_t$, only a single pair of simultaneous states $(U_t, V_t)$ will exist. On the other hand,  a reconstructed state $U_t$ in $R(U)$ corresponds to a non-singleton set of points $T_{V,U}^{-1} T_{U,U}\, U_t$ in $R(V)$, since $T_{V,U}$ is not one-to-one. This is the asymmetry referred to earlier. For each $U_t$, there is a multiple point set of simultaneous pairs $(U_t,V_t)$, unlike the case for $V_t$. If we can use the observed time series to distinguish betwen the two cases, one point versus multiple points, for pairs $(U_t,V_t)$, we can infer in which direction the coupling exists. 

The case 2 of bidirectional coupling is simpler, because under generic conditions, each of the reconstructions $R(U)$ and $R(V)$ are in one-to-one correspondence with states of the bidirectionally-coupled system $U\leftrightarrow V$, and therefore $R(U)$ and $R(V)$ are in one-to-one correspondence. Unlike the asymmetric case (1), in case (2), for each $U_t$ in $R(U)$ there is a unique $V_t$ in $R(V)$, and vice versa.

Now consider the case 3 of latent coupling, where there is an unobserved driver $D$ that is coupled to both the observed systems $U$ and $V$, but no direct connection between $U$ and $V$, as in the diagram:
\begin{center}
\begin{tikzcd}[column sep=small]
& D \arrow[dl] \arrow[dr] & \\
U & & V
\end{tikzcd}
\end{center}

Then, generically, observations from $U$ (resp. $V$) reconstruct the skew product $D\to U$ (resp., $D\to V$). This results in the diagram
\begin{center}
\begin{tikzcd}[column sep=small]
& R(U)\arrow[dl,swap,"T_{U,U}"] \arrow[dr,swap,"T_{U,D}"] & &R(V)\arrow[dl,"T_{V,D}"] \arrow[dr,"T_{V,V}"]& \\
U  & & D & & V 
\end{tikzcd}
\end{center}
where none of the four maps are one-to-one. Now both $T_{U,D}^{-1} T_{V,D}\, V_t$ and $T_{V,D}^{-1} T_{U,D}\, U_t$
 comprise sets of multiple points, i.e. more than one. This observation suggests a criterion for $U$ and $V$ being driven by an unobserved system, which is that for each $U_t$, there are multiple simultaneous $(U_t, V_t)$, and the same for each $V_t$.
 
 Although not essential for the goals of the present article, this structure can be exploited to theoretically reconstruct a latent driver $D$ from time series observations as follows. Consider a state $V_t$ in $R(V)$. In this case, $T_{U,D}^{-1} T_{V,D}\, V_t$ consists of multiple points $\{U_i\}$ by assumption, usually dispersed along a proper subset of $R(U)$. For each of the $U_i$, we can repeat the process in the opposite direction, which represent different indices in the time series.  By assumption, the $T_{V,D}^{-1} T_{U,D}\, U_i$ are multiple points sets in $V$, only one of which is the original $V_t$. Recursing this process identifies an equivalence class of point sets in both $R(U)$ and $R(V)$, which corresponds to a theoretical state $d$ linked to the multiple points in $R(U)$ and $R(V)$. If we repeat this construction for all $V_t$ in $R(V)$ and $U_t$ in $R(U)$, these equivalence classes represent the states of an unobserved attractor $D$ that implies scenario 3, i.e. latent coupling $D\to U$ and $V$. This process of reconstructing the driver was the subject of \cite{sauer2004reconstruction}, which was recently put on a more stable numerical  footing \cite{gilpin2025recurrences}.

The preceding discussion clarifies the critical distinction between the cases 1 - 3, which depends on whether the sets $(U_t, V_t)$, for fixed $U_t$ or $V_t$, are singleton sets or more than one element. The possibilities can be efficiently summarized in a diagram:

\medskip

\begin{tabular}{c||c|c||}
\multicolumn{1}{c}{} & \multicolumn{1}{c}{($U_t$, many $V_s$)} & \multicolumn{1}{c}{($U_t$, single $V_s$)}  \\
\hline
(many $U_s$,\ $V_t$)&\begin{tikzcd}[column sep=small]
& D \arrow[dl] \arrow[dr] & \\
U & & V
\end{tikzcd}
& $V\to U$\\
\hline
(single $U_s,\  V_t$) & $U\to V$ &  \begin{minipage}{1in} \begin{center}\vspace*{.1in} $U\leftrightarrow V$\\ or\\ Generalized Synchrony \vspace*{.1in} \end{center}
\end{minipage}\\
\hline
\end{tabular}

\bigskip

For example, the diagram says that on the reconstructed combined system attractor $R(U)\times R(V)$, if each $U_t$ belongs to a set $(U_t, V_s)$ with several different $V_s$, and if each $V_t$ belongs only to a single $(U_s, V_t)$, this is evidence for the coupling $U\to V$. Alternatively, if each $U_t$ belongs to a set $(U_t, V_s)$ with several different $V_s$, and if each $V_t$ belongs to a set $(U_s, V_t)$ with several different $U_s$, then evidence exists for a latent coupling by a third, unobserved system. While this is a compelling theoretical description, it remains to propose a way to determine which of these cases are likely to have produced the time series under study.

The following statistical test is designed to determine which of these cases hold, using only the observed time series. The logic is as follows: If for a given $V_t$, there is a unique corresponding time delay vector $U_t$, then if we look at the $U_s$ that are time-contemporaneous with $V_s$ that are close to $V_t$, these $U_s$ should be close to $U_t$. The test checks to see whether the $U_s$ progressively move farther from $U_t$ as the $V_s$ are moving away from $V_t$.

\medskip

\noindent {\bf Direction of Coupling Test: DirC(U,V).} For each $t$ and $1\le j \le n$, find the $j$th nearest neighbor of the delay vector $V_t$ and denote its time $s_j$. Sort the entire set of distances $|U_t - U_s|$ for $1\le s\ne t \le N$, and find the percentile rank $r_{(j)}$ of the distance $|U_t - U_{s_j}|$ among the entire set. Let $\hat{r}_{(j)}$ denote the average percentile rank over all $t$. Plot the best fit line through the points $(j,\hat{r}_{(j)})$.

A slope greater than zero is evidence that for each $V_t$, the set of possible simultaneous pairs $(U_t,V_t)$ is a single pair. A slope indistinguishable from zero is evidence that it is a set of multiple pairs (i.e., not a singleton).

\medskip

Confidence intervals for the DirC test can be developed similarly to the DetC test. For the line fit $y = \alpha+\beta t$ where we use a one-sample Student-t test to compare the slope $\beta$ with the null hypothesis $\beta = 0$, the radius of the $95\%$ confidence interval is $s = (2\sum_{i=1}^n (y_i - \hat{y}_i)^2)/((n-2)\sum_{i=1}^n (x_i-\overline{x})^2)$.

Figure \ref{f1}(b) demonstrates the use of DirC. Here the two 2-cell H\'enon maps $U$ and $V$ are coupled as $U\to V$.  The slope in DirC(U,V) is positive, which provides evidence that for each $V_t$, there is a unique $U_t$ that can exist simultaneously with $V_t$. On the other hand, the slope of DirC(V,U) cannot be distinguished from zero, meaning that there are multiple $V_t$ corresponding to a given $U_t$. These results are consistent with case 1, unidirectional coupling from $U$ to $V$.

\begin{figure}
\subfigure[]{\includegraphics[width=.23\textwidth]{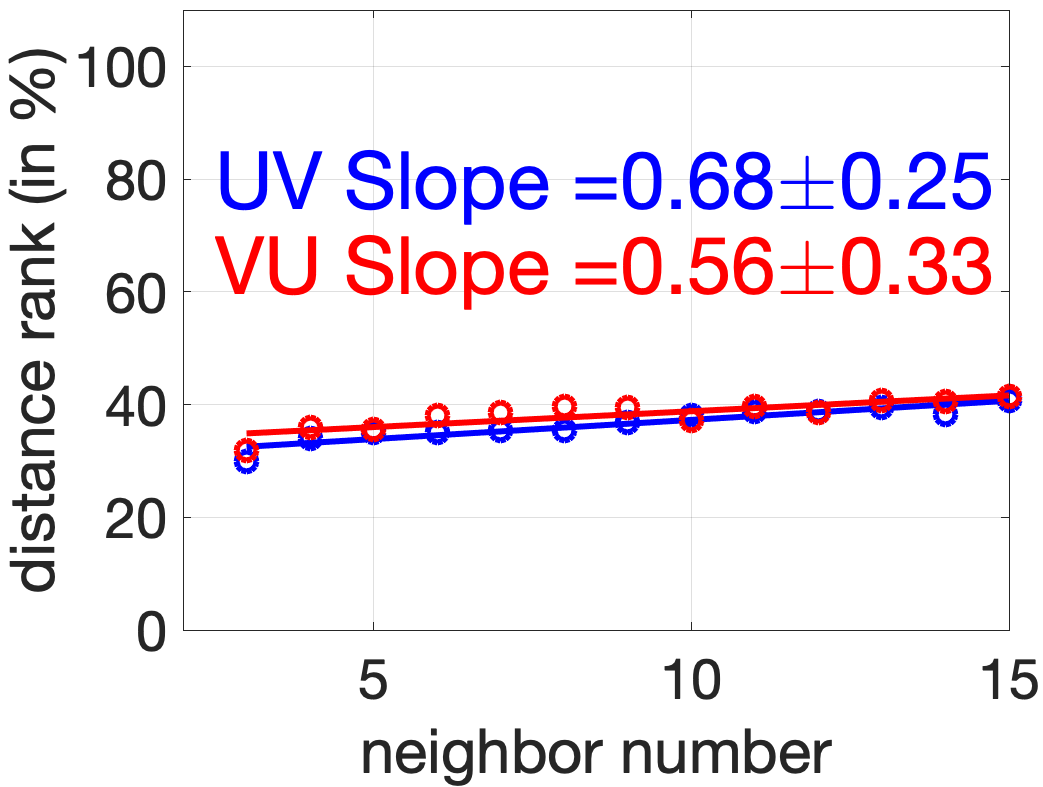}}\hspace*{.02in}
\subfigure[]{\includegraphics[width=.23\textwidth]{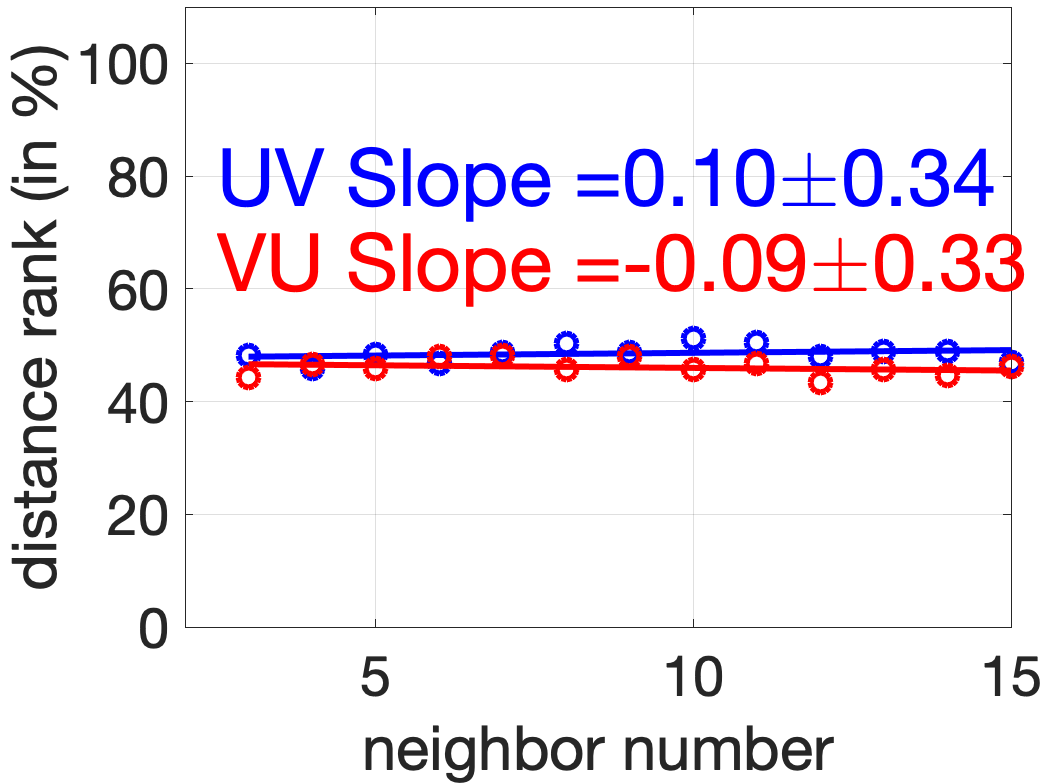}}
\caption{\label{f2} DirC test applied to  (a) bidirectional coupling $U\leftrightarrow V$ and (b) latent coupling $D\to U, D\to V$. DirC(U,V) in blue and DirC(V,U) in red. (a) Both slope estimates reject zero slope, correctly implying bidirectional driving.  (b) Neither of the slope estimates reject the null hypothesis of zero slope, correctly implying latent coupling from an unobserved common driver. }
\end{figure}

The two tests proposed in this article can be used together in the following way. If the Detection of Coupling Test is passed in either of the $U$ or $V$ directions, the remaining possible relations are 1 - 3. Then, the Direction of Coupling Test applied to both $U$ and $V$ will distinguish between the remaining four possibilities. In the table, DirC(U,V) + means the slope is determined to be greater than zero, and DirC(U,V) 0 means zero slope cannot be rejected.

\bigskip

\begin{tabular}{||c||c|c||}
\multicolumn{1}{c}{} & \multicolumn{1}{c}{DirC(V,U) $0$} & \multicolumn{1}{c}{DirC(V,U) $+$}  \\
\hline
\hline && \\
DirC(U,V) $0$ & \ latent $D\to U,V$\ \ \ & $V\to U$\ \ \ \\ && \\
\hline && \\
DirC(U,V) $+$ & $U\to V$ &\ $U\leftrightarrow V$ or GS\ \ \ \\  && \\
\hline   
\end{tabular}

\bigskip

Fig.~\ref{f2} demonstrates the use of the DirC test for time series collected from two 2-cell H\'enon maps in cases 2 and 3. In Fig.~\ref{f2}(a), the times series are generated from a system $U\leftrightarrow V$ with two-way driving. The DirC(U,V) and DirC(V,U) both identify positive slope outside the confidence interval, concluding bidirectional driving as in case 2. In Fig.~\ref{f2}(b), zero slope cannot be rejected in either DirC(U,V) or DirC(V,U), indicating latent driving from an unobserved system as in case 3.

The tests can also be applied to continuous systems such as coupled Lorenz \cite{lorenz1963deterministic} systems. For example, let $U, V$, and $W$ each be systems of the form
\begin{eqnarray} \label{e2}
\dot{x} &=& \sigma (y-x) \nonumber\\
\dot{y} &=& -xz+\rho x - y\\
\dot{z} &=& xy - \beta z\nonumber
\end{eqnarray}
with slightly different parameter values near $\sigma = 10, \rho = 28, \beta = 8/3$. Further, we couple them in a small network as in (A) or (B) below,
\begin{center}
(A)\begin{tikzcd}[column sep=small]
& U \arrow[dl] & \\
V \arrow{rr} & & W \arrow[ul]
\end{tikzcd} \hspace{.3in}
(B)\begin{tikzcd}[column sep=small]
& U \arrow[dl] \arrow[dr] & \\
V \arrow{rr} & & W
\end{tikzcd}
\end{center}
where each arrow represents a term $cx$ proportional to the $x$-variable of the driving Lorenz system added to the $y$-variable of the target system.  The following table shows the results of applying DirC to both scenarios, using $c = 5$:

\noindent
\begin{tabular}{||c|c|c||}
\multicolumn{1}{c}{} & \multicolumn{1}{c}{(A)} & \multicolumn{1}{c}{}  \\
\multicolumn{1}{c}{} & \multicolumn{1}{c}{slope} & \multicolumn{1}{c}{}  \\
\hline
$U\to V$ & $\ 0.68 \pm 0.13\ $ & $+$\\
\hline
$V\to U$ & $\ 0.50 \pm 0.26$ & $+$\\
\hline
$V\to W$ & $\ 0.61 \pm 0.17\ $ & $+$\\
\hline
$W\to V$ & $\ 0.39 \pm 0.22$ & $+$\\
\hline
$W\to U$ & $\ 0.62 \pm 0.24\ $ & $+$\\
\hline
$U\to W$ & $\ 0.42 \pm 0.14$ & $+$\\
\hline
\end{tabular}\hspace*{.15in}
\begin{tabular}{||c|c|c||}
\multicolumn{1}{c}{} & \multicolumn{1}{c}{(B)} & \multicolumn{1}{c}{}  \\
\multicolumn{1}{c}{} & \multicolumn{1}{c}{slope} & \multicolumn{1}{c}{}  \\
\hline
$U\to V$ & $\ 0.76 \pm 0.20\ $ & $+$\\
\hline
$V\to U$ & $\ 0.15 \pm 0.36$ & $0$\\
\hline
$V\to W$ & $\ 0.25 \pm 0.17\ $ & $+$\\
\hline
$W\to V$ & $\ 0.13 \pm 0.16$ & $0$\\
\hline
$W\to U$ & $\ 0.07 \pm 0.23\ $ & $0$\\
\hline
$U\to W$ & $\ 0.47 \pm 0.22$ & $+$\\
\hline
\end{tabular}

\medskip

The slopes are shown along with the 95\% confidence intervals. A plus sign in the right column means that the corresponding coupling was determined to exist by the test, and $0$ means the coupling was not detected to the confidence level.  There are six possible couplings. We note that in scenario (A), all three systems are driven by each of the others, directly or indirectly, and accordingly, the DirC test detects that coupling in each case. In scenario (B), the DirC test correctly detects the three drivings $U\to V,\ V\to W$, and $U\to W$, and finds no evidence of the other three possibilities.

\begin{figure}
\subfigure[]{\includegraphics[width=.23\textwidth]{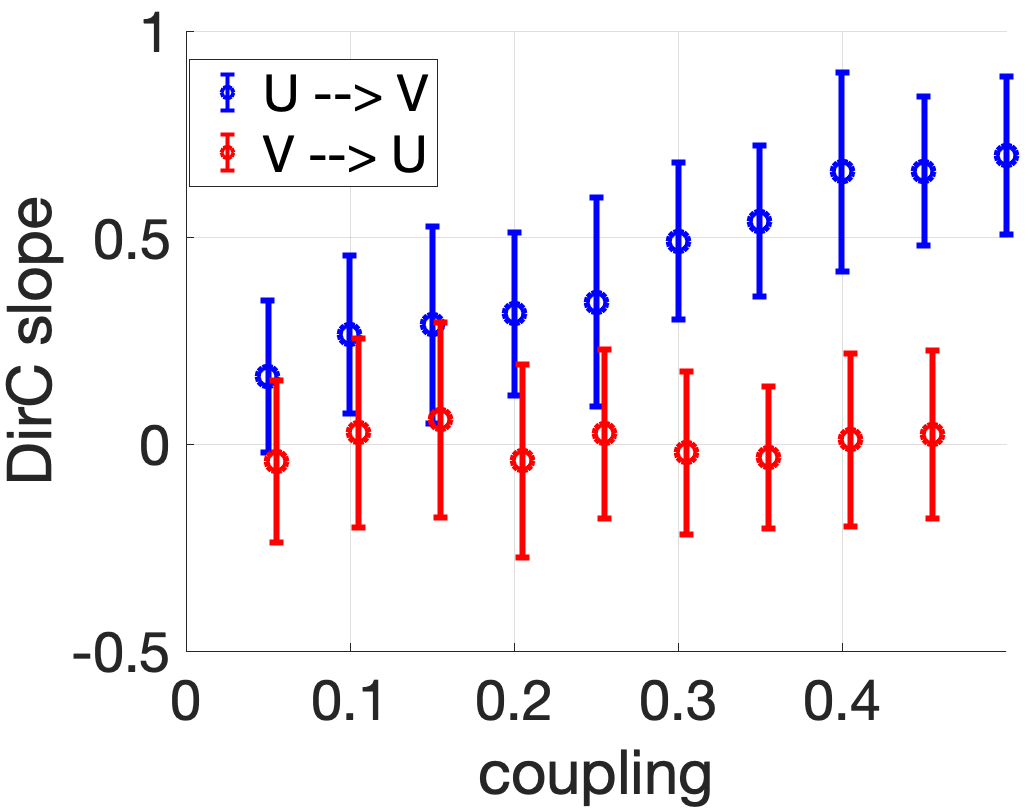}}\hspace*{.02in}
\subfigure[]{\includegraphics[width=.23\textwidth]{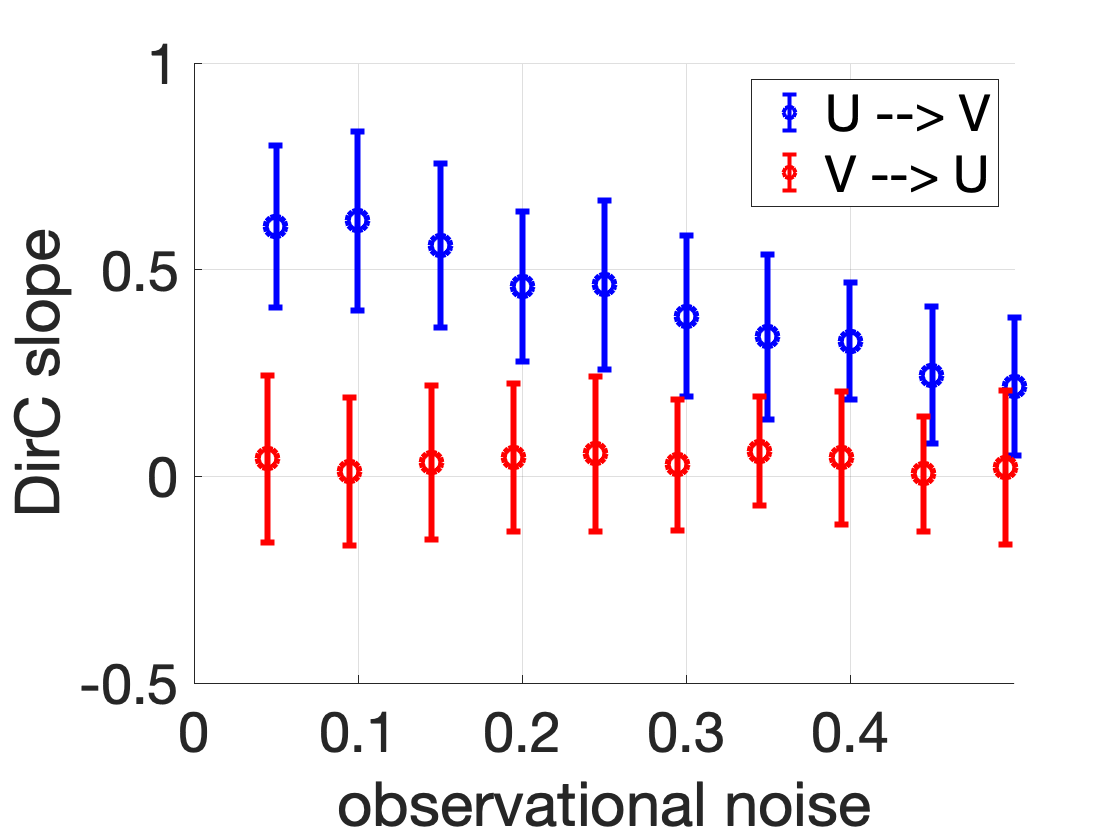}}
\caption{\label{f7} Dependence  of DirC statistic on coupling strength and noise for unidirectional driving $U\to V$. (a) The mean of the DirC slope over 50 realizations of length 1000 times series of $U\to V$ for two-cell H\'enon examples is plotted versus coupling strength along with its standard error.  (b) DirC slope plotted versus observational noise. }
\end{figure}

To investigate the sensitivity of the DirC test with respect to parameters, we compared the results for different coupling strengths and observational noise levels. Fig.~\ref{f7}(a) shows the dependence on the DirC test on coupling strength. The data series of length 1000 was generated from a coupled system $U\to V$ with dynamics from the two-cell H\'enon map (\ref{e8}). The coupling between the two networks is achieved by adding a constant $c$ times the $x$-variable from $U$ to the $u$-variable of $V$. Fig.~\ref{f7}(a) plots the DirC statistic versus the coupling strength $c$. For small $c$, the  $95\%$ confidence intervals of statistics DirC(U,V) and Dir(V,U) overlap. For increased $c$, they no longer overlap and  definitively pick up the correct direction of driving. 
Fig.~\ref{f7}(b) shows that the DirC test degrades gracefully with increased observational noise. For small noise the test easily identifies the driving $U\to V$, and as noise is increased, the test begins to fail.

\section{Generalized synchrony} \label{sGS}

As mentioned above, the phenomenon of generalized synchrony plays a clarifying role in the study of causation between dynamical systems. A typical example is given by a coupled skew system $U\to V$ where there is no feedback from $V$ to $U$, but where a one-to-one correspondence develops between states of $U$ and states of $V$. That is, after transient behavior, each state of $U$ coexists with a unique state of $V$, and vice versa. This can occur even when $U$ and $V$ are different dynamical systems. 

On the one hand, by its existence, GS shows that Granger causality cannot detect the fact that $U\to V$, since knowledge of the $U$ state cannot add to the ability to predict future states of $V$. (In fact, the logic of such an ability would be circular, since one could say the same in the opposite direction.) Furthermore, GS is also a wild card for the DirC calculation, because the sets of possible pairs $(U_t,V_t)$ discussed in the derivation of the DirC method are singletons, both in the non-GS case of bidirectional coupling $U\leftrightarrow V$ and the case of GS caused by unidirectional coupling or by latent coupling. In other words, assuming there is no time-lag in the functional coupling of $U\to V$, for deterministic dynamical systems it is challenging for any method to distinguish GS under $U\to V$ from the relationship $U\leftrightarrow V$ on the basis of observed time series alone.

\begin{figure}
\subfigure[]{\includegraphics[width=.23\textwidth]{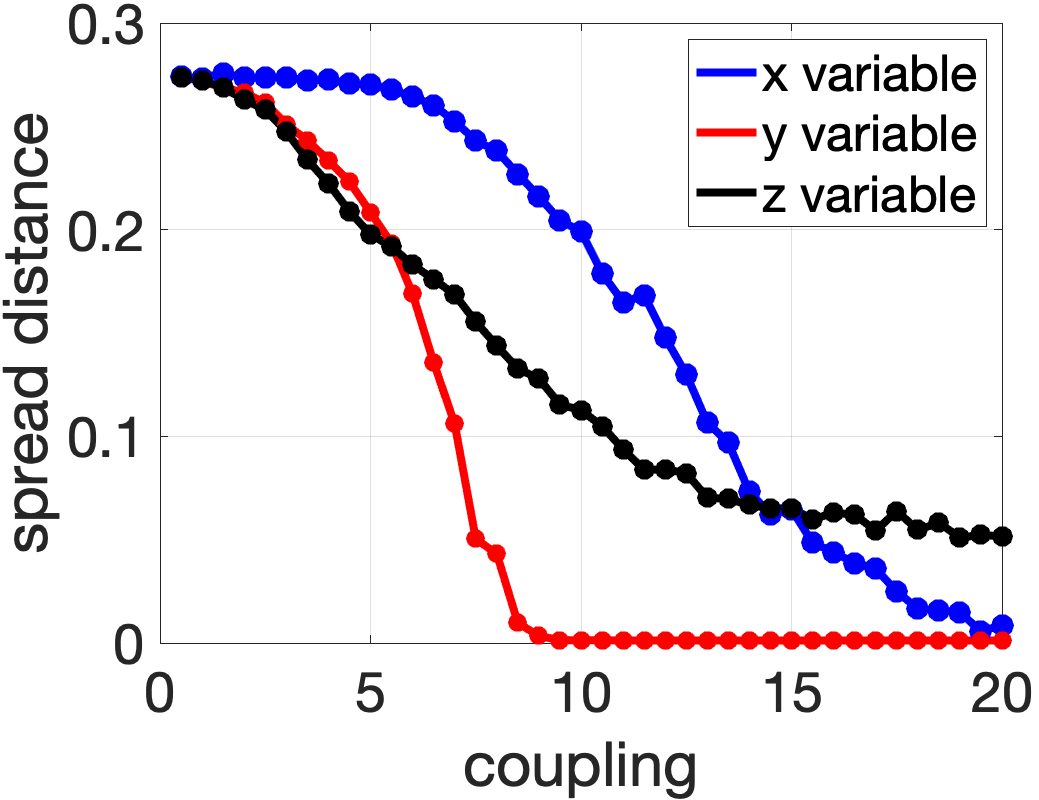}}\hspace*{.02in}
\subfigure[]{\includegraphics[width=.23\textwidth]{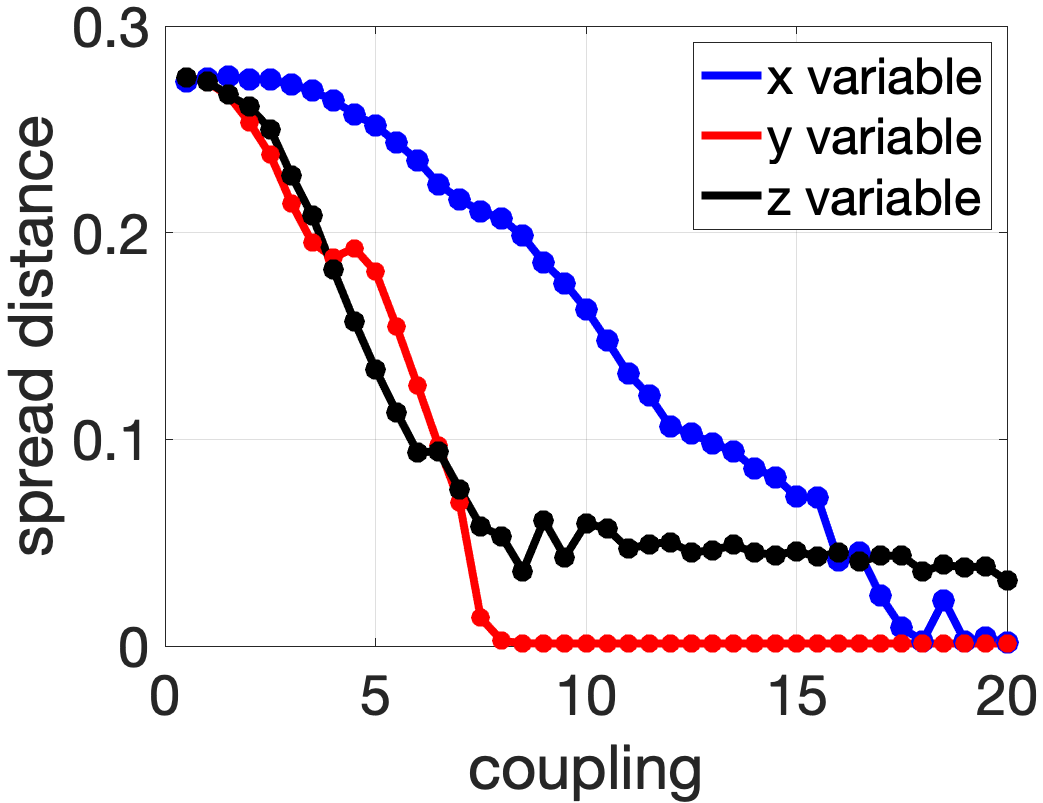}}
\caption{\label{f4} Generalized synchrony from coupling $U\to V$. The plot shows the average spread of states of $V$ chosen contemporaneously with a state of $U$. GS is indicated when the spread drops to near zero. (a) Coupling $cx$ from Lorenz $U$ to $x,y$ or $z$-variable of Lorenz $V$. GS occurs for coupling greater than $0.1$ for the $y$-variable and about $0.25$ for the $x$-variable. Driving the $z$-variable does not achieve GS in this range of driving. (b) Coupling $cx$ from R\"ossler $U$ to $x,y$ or $z$-variable of Lorenz $V$. The RMS spread of a set of $k = 12$ states was used in these plots.}
\end{figure}

We include here two illuminating examples of generalized synchronization for chaotic flows. The first is  constructed from two Lorenz systems with different parameters. Let $U$ and $V$ be systems of form (\ref{e2}) with parameters $\rho = 27$ and $\rho = 30$, respectively. Figure \ref{f4}(a) shows that with coupling of $5x$ from $U$ added to the $y$-variable of $V$, the system is not in GS. However, changing the coupling to $15x$  induces GS between $U$ and $V$. On the other hand, coupling of $cx$ from $U$ to the $x$-variable of $V$ requires $c>20$ to achieve GS.

 To positively verify GS in this example, we assume knowledge of all six phase variables of $U$ and $V$. A number $k$ of initial conditions $(u,v_1),\ldots, (u,v_k)$, where the $v_i$ are randomly chosen, are used to create trajectories of the coupled system $U\to V$.  After deleting the transients, the $k$ trajectories represent states of the skew attractor $U\times V$ with identical $x$-state.  Then we calculate the standard deviation of the set of distances of the $k$ states of $V$ from their center of mass. The signature of global synchrony is when this standard deviation, or spread, of points drops to near zero.
 
 The mean of this spread, normalized by the size of the attractor, as we sweep over states on $U$ is plotted in Figure \ref{f4}(a). For small coupling $U\to V$ achieved by adding a coupling strength times the $x$-variable of $U$ to the $y$-variable of $V$, there is no synchronization. When coupling strength increases beyond $10$, the systems are in generalized synchrony. The same is true for adding the $x$-variable of $U$ to the $x$-variable of $V$, for coupling strength beyond 20. On the other hand, Figure \ref{f4}(a) also shows that using a coupling proportional to $x$ added  to the $z$-variable of $V$ does not lead to GS in the range of coupling strengths shown.

\begin{figure}
\subfigure[]{\includegraphics[width=.23\textwidth]{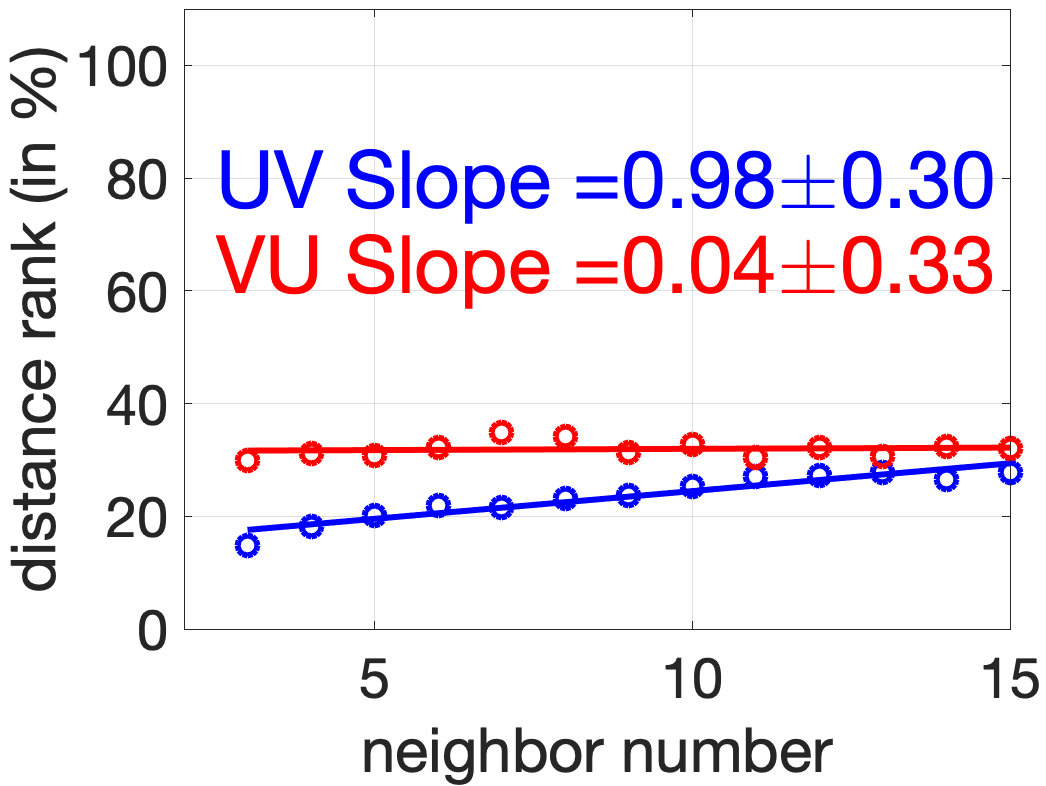}}\hspace*{.02in}
\subfigure[]{\includegraphics[width=.23\textwidth]{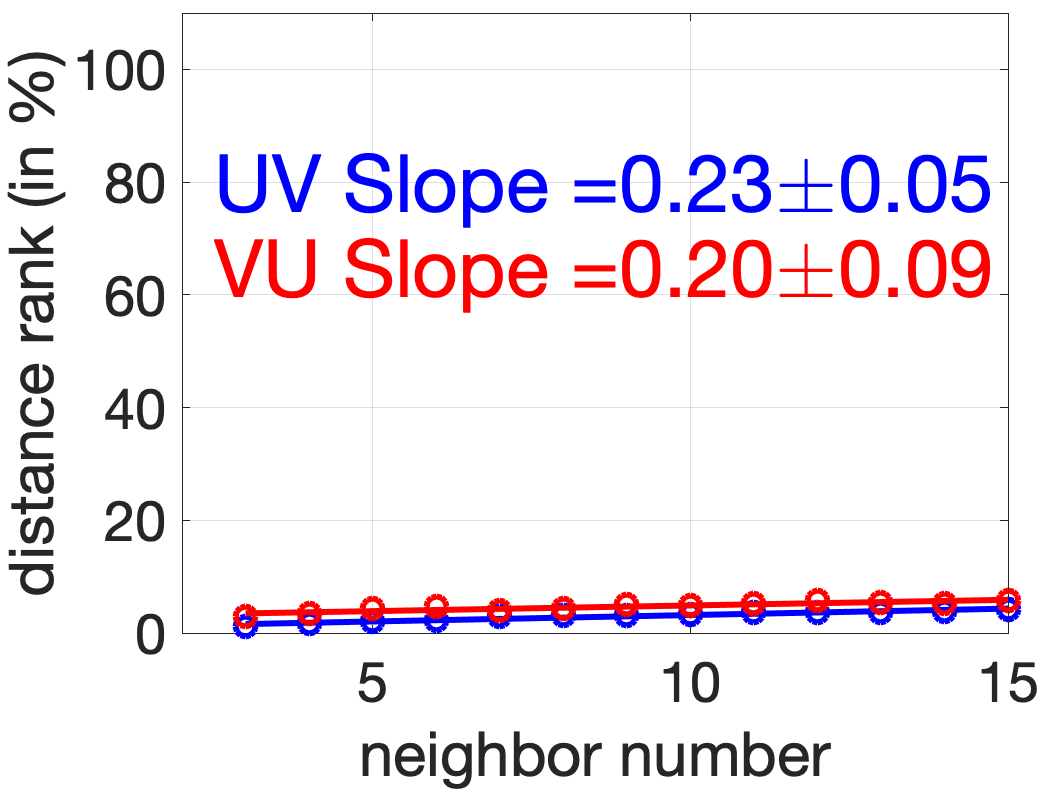}}
\caption{\label{f5} DirC statistic for data from Lorenz systems $U\to V$, DirC(U,V) in blue and DirC(V,U) in red.   (a) For coupling $c = 5x$ added to $y$-variable of $V$ vector field, (not in general synchronization),  only UV slope statistically greater than zero, implying case (1), $U\to V$. (b) Coupling $c = 15x$ (forces generalized synchronization), UV and VU slopes significantly greater than zero implies case (2), $U\leftrightarrow V$ or GS. Data consists of the $x$-coordinate time series of 1000 points observed from $U$ and $V$ with sampling interval $\Delta t = 0.1$. }
\end{figure}

 When we can only measure time series of $U$ and $V$, and have a smaller data set, we can apply the DirC test to distinguish between unidirectional coupling $U\to V$ that does and does not result in GS. The results of analyzing a 1000 point time series of $x$-coordinates observed from $U$ and $V$ is shown in Fig.~\ref{f5}.
 In
 Figure \ref{f5}(a), $U$ drives $V$ by adding $5x$ to the $y$-variable of $V$. Fig.~\ref{f4}(a) indicates that generalized synchrony does not occur in this case. Accordingly, DirC  correctly concludes unidirectional coupling $U\to V$.  
 When the driving is increased to $15x$, generalized synchrony occurs,  and we find that DirC rejects zero slope in both directions, as shown in Fig.~\ref{f5}(b). The explanation for this result is generalized synchronization caused by unidirectional coupling $U\to V$, not two-way driving $U \leftrightarrow V$.

For a Lorenz attractor driven by the output of another Lorenz attractor, GS is relatively easy to achieve, even for non-identical systems as above. If the driving system sends a signal $cx$ to the $x$ or $y$ variable of the response system for $c$ sufficiently large (as in the above paragraph), the system enters GS. 

The second example of generalized synchrony occurs for the coupling $U\to V$ where $U$ is a R\"ossler attractor driving a Lorenz attractor $V$, with the standard parameters $\sigma = 10, \rho = 28, \beta = 8/3$. The R\"ossler equations \cite{rossler1976equation} are
\begin{eqnarray} \label{e4}
\dot{x} &=& -y - z \nonumber\\
\dot{y} &=& x+ay\\
\dot{z} &=& b+(x-d)z\nonumber
\end{eqnarray}
and we will use the parameter settings $a = 0.1, b = 0.1$, and $d = 14$. As before,  the Lorenz attractor $V$ is driven by adding a signal $cx$ from the $U$ system to the $x, y$ or $z$ variable of $V$.
Fig.~\ref{f4}(b) shows that moderately-sized couplings from the $x$ variable of the R\"ossler $U$ to the $x$ or $y$ variable of Lorenz $V$ also cause generalized synchrony.

\section{Discussion}

The tests proposed here are designed to learn the coupling characteristics of a pair of time series, produced simultaneously by two different deterministic processes. Our goal is to exploit the fact of asymmetry between the dual reconstructions as simply and robustly as possible, in order to tease conclusions from time series that may be short and perturbed by observational noise. These tests may be applied to univariate time series (or multivariate time series) measured from two or more processes,  as long as the state space reconstructions afford a one-to-one correspondence to the original systems.

Among the innovations of our methods are the reduction of the information of distances between points by treating them ordinally, in order to make the results robust to measurement noise. This also allows us to apply the theory of order statistics in the DetC test to determine dependence of time series, and to retrieve statistical conclusions by assigning confidence intervals. The DirC test exploits the asymmetry of paired Takens delay coordinate reconstructions in order to distinguish unidirectional, bidirectional, and latent coupling. Our goal in this article is to deploy these innovations in the simplest possible way so that they may succeed with minimal data requirements.

These tests will gradually fail in general for stochastic systems, due to the degradation of the delay coordinate embedding that is foundational for this approach. Other methods more suited toward working under stochastic assumptions will work better in nondeterministic contexts, although where the trade-off occurs will be dependent on the details of the situation.

A key requirement for faithfulness of the delay coordinate embedding is genericity of the dynamics and observations. This includes both the internal dynamics of $U$ and $V$, and in addition, the connections (if any) between the two systems. We can expect this genericity to exist normally in natural systems, in the absence of a particular structural reason that defeats the hypotheses of Takens' theorem. However, we are mindful that the generic case is in a sense the simplest base case, and that when non-generic structural elements occur in the systems under study, the methods proposed here will need to be refined accordingly.

The strict hypotheses of Takens' theorem refer to systems being observed simultaneously. However, the tests developed here may be of use in systems with a time lag by making the modeling hypothesis, for example, that a time lagged system $U_\tau$ is driving the system $V$. Coupling established in this way can be used to confirm directionality for systems that communicate with time lags, but otherwise might be difficult to represent by finite-dimensional attractors. 

The existence of generalized synchrony presents an interesting complication to the problem of distinguishing coupling direction from time series. On the one hand, when couplings exist in both directions, i.e. $U \leftrightarrow V$, under generic conditions, the delay coordinate embedding from either $U$ or $V$ will uniquely reconstruct a corresponding state from the other that will occur simultaneously, so that a one-to-one correspondence exists between reconstructed states. This is also the definition of generalized synchrony, but as demonstrated here, GS can occur solely from unidirectional coupling $U\to V$. Therefore evidence such as the DirC results in Fig.~\ref{f5} cannot distinguish between these two cases.

\acknowledgments 
Research partially supported by NSF DEB-1655203, NSF ABI-1667584, Department of Interior-NPS-P20AC00527, the McQuown Fund and the McQuown Chair in Natural Sciences, University of California, San Diego. We thank the anoonymous reviewers for significantly improving the manuscript.

\bibliography{drive}

\end{document}